\documentclass[12pt]{iopart}

\usepackage{graphics}
\begin{document}

\title[Nuclear electron capture rate in $^7$Be]{Nuclear electron capture rate in stellar interiors and the case of $^7$Be}
\author{P. Quarati$^{1,2}$ and A.M. Scarfone$^{1,3}$}
\address{$^1$Dipartimento
di Fisica - Politecnico di Torino, Italy.\\
$^2$Istituto
Nazionale di Fisica Nucleare (INFN), Sezione di Cagliari, Monserrato, Italy.\\
$^3$Istituto
Nazionale di Fisica della Materia (CNR-INFM), Unit\'a del Politecnico di Torino, Italy.}
\eads{\mailto{piero.quarati@polito.it}\\ \mailto{antonio.scarfone@polito.it}}

\date{\today}
\begin {abstract}
Nuclear electron capture rate from continuum in an astrophysical plasma environment (like solar core) is calculated using a modified Debye-H\"{u}ckel screening potential and the related non-Gaussian $q$-distribution of electron momenta. For $q=1$ the well-known Debye-H\"{u}ckel results are recovered. The value of $q$ can be derived from the fluctuation of number of particles and temperature inside the Debye sphere. For $^7$Be continuum electron capture in solar core, we find an increase of $7$ -- $10$ percent over the rate calculated with standard Debye-H\"{u}ckel potential. The consequence of this results is a reduction of the same percentage of the SSM $^8$B solar neutrino flux, leaving unchanged the SSM $^7$Be flux.
\end {abstract}
\pacs{23.40.-s, 97.10.Cv, 52.20.-j, 05.20.-y}
\submitto{\JPG}
\maketitle

\section{Introduction}

Since the early works by Bethe \cite{Bethe} and Bahcall \cite{Bahcall1,Bahcall2} great attention has been devoted to the screening effect of Coulomb potential on electron capture (EC) by nuclei in astrophysical plasmas \cite{Clayton} and on its implications with neutrino production.\\
The plasma influence has been in great part explored by means of Debye-H\"{u}ckel (DH) treatment. However, Johnson et al. \cite{Johnson} found that assumptions for the validity of DH potential are strongly violated in stellar cores, because, among other reasons, the requirement to have many particles in a Debye sphere is not fulfilled. At solar conditions there are only few (about four) particles per Debye sphere. R\^{o}le and importance of several effects in the electron component of a weakly nonideal hydrogen plasma have been recently investigated in \cite{Starostin}, inducing us to investigate on EC beyond DH approach.\\
The contribution to nuclear EC rate of bound and continuum electrons has been widely studied by Shaviv and Shaviv \cite{Shaviv}. They have found that under solar conditions all elements with
$4<Z<12$ are fully ionized, therefore the contribution from bound electrons to the capture rate can be disregarded. Formerly, Bahcall and Moeller \cite{Bahcall3} reported that the ratio of the rate from continuum over the rate from bound states is about $1.2$. \\
Fluctuations of electric microfields and Debye particle numbers have been investigated by Gruzinov and Bahcall \cite{Gruzinov} and by Brown and Sawyer \cite{Brown} within the Feynman density matrix treatment and the mean field theory applied in the Boltzmann limit validity, i.e. at global thermodynamical equilibrium. They have found that non-spherical fluctuations can change the reaction rate of about $1$ -- $2$ percent and that quantum corrections to Maxwell-Boltzmann (MB) limit for $Z=4$ are less or about $1$ percent.\\
Screening of Coulomb field of the nucleus by outer electrons has been explored by considering the Hulth\'{e}n potential that has a shape very close to a screened Coulomb potential. By solving analytically Klein-Gordon and/or Schr\"oedinger equations one can calculate the electron density at the nucleus and consequently derive the Fermi factors in terms of hypergeometrical functions \cite{Durand}.\\
In all above quoted papers, continuum EC rates are evaluated in the classical statistics limit, using the MB electron momentum distribution that is the correct distribution for Coulomb and DH potential $V_{_{\rm DH}}(r)$, but is not the appropriate distribution to be used with the modified DH (MDH) potential $V_q(r)$ ($V_{q=1}(r)=V_{_{\rm DH}}(r)$) introduced by us in \cite{Quarati}. The spatial charge distribution related to the last mentioned screening potential differs from the others having a spatial cutoff. The linear Poisson equation used to deduce the screening DH potential must be substituted by a non-linear equation to take into account particle correlations and fluctuations and $V_q(r)$ is deduced as a power law function.
In \cite{Quarati} we have also derived the potential $V_q(r)$ in the framework of the super-statistics approach \cite{Beck,Wilk}, by considering an inverse Debye radius fluctuating around its average value given by $\xi=\langle1/R_{\rm DH}\rangle^{-1}$. The evaluation of the rate requires non-Gaussian generalised $q$-distribution of electron momenta with the same value of $q$ of the spatial charge density distribution.\\ In this work, we evaluate the variation of the EC rate due to the MDH screening potential over the pure Coulomb screening rate as a function of average Debye radius $\xi$ and entropic non-extensive parameter $q$ \cite{Tsallis,Haubold,Utyuzh}. We consider the case $0<q<1$ to have a screening potential with finite spatial range. This peculiarity of some high density astrophysical plasmas and of most of the Sun is due to the fact that, in these systems, the mean interparticle distance is smaller or slightly smaller than DH radius \cite{Shaviv}. Standard DH results are obtained in the limit $q\to1$ of the electron non-Gaussian distribution.\\
The value of the parameter $q$ can be fixed by means of its relation with the fluctuation of particles number and of temperature inside the Debye sphere, therefore the same value of $q$ can be used both for spatial charge density and for momentum distributions.\\
When this treatment is applied to EC by $^7$Be in the solar core,
we can derive the amount of the rate enhancement for the electrons of continuum over the DH rate. We find that, at solar conditions, the rate can increase of about $7$ -- $10$ percent over DH screening rate with $q$ ranging between $0.84$ and $0.88$ (small deformations of MB distribution).
This result can be useful in the interpretation of the observed $^7$Be and $^8$B neutrino fluxes, in the evaluation of relevant astrophysical $S$ factors and of CNO solar neutrinos \cite{Borexino,Brown1,Stonehill,Haxton}. For recent discussions and comments on the EC by $^7$Be from a continuum three-body initial state we send to \cite{Belyaev2} and references therein. See \cite{Chakrabarty} for EC rate evaluated with a percentage of non-thermal fat-tail electron distribution.\\
In Section 2, we report the expressions of the rates evaluated with $q$-generalized distribution at few different values of $q$. In Section 3, we present numerical results of the rates, for electrons belonging to the continuum for different values of the parameter $q$ and focusing to the case of EC by $^7$Be in the solar core. Finally, in Section 4, we outline our conclusions.

\section{Rates}

It is known that in nuclear continuum EC the evaluation of the rates includes the Fermi factor, i.e. the electronic density at the limit of $r\to0$ in a pure Coulomb potential. Therefore, the pure nuclear rate is corrected because of the Coulomb interaction between the captured electron and the nucleus. Usually, within a plasma environment, DH potential is adopted in place of Coulomb potential. In this case the electron density at the nucleus is known only numerically from the solution of the appropriate Schr\"oedinger equation.\\
In an astrophysical environment, the momentum distribution of  screening electrons can differ significatively from MB distribution. We have recently introduced a MDH screening potential to take into account non-linear and correlation effects by means of fluctuation of $1/\xi=\langle1/R_{\rm DH}\rangle$ \cite{Quarati}. Of course, also in this case we need to evaluate the electron density at $r=0$ by solving the appropriate Schr\"odinger problem for electrons in the continuum.\\
Whereas Schr\"oedinger solution with a Hulth\'{e}n potential can be given in a close analytic form,
the MDH potential (as well as the standard DH potential) admits only numerical solutions for the electron density at $r=0$. Alternatively, one can use the Hulth\'{e}n potential that fits quite well the MDH potential in the small $r$ region (near the nucleus), but contains an infinite tail instead of having a cut-off at an appropriate value of $r\equiv r_{\rm cut}$ as for the MDH potential with $q<1$. A Coulomb cut-off potential, obtained by imposing the condition $V_{_{\rm C}}(r)=0$ for $r>r_{\rm cut}$ and consequent discontinuity in the potential at $r=r_{\rm cut}$, was used in the past as the simplest way to screen the Coulomb field \cite{Ford}. Otherwise, in our approach, the cut-off condition arises naturally and the MDH potential is a smooth function for any $r>0$, vanishing for $r>r_{\rm cut}$.

In the following, we evaluate the rate for the free electron capture by a $(A,\,Z)$ nucleus, given by the integral, in the three dimensional space of velocities, of electron capture cross-section $\sigma_{\rm e}$ times the electron velocity $v$, the normalised probability density $(F_{_{\rm C}},\,F_{_{\rm DH}},\,F_{_{\rm H}}$ or $F_q)$ that an electron of the continuum spectrum, with velocity $v$ and travelling in a screening potential $(V_{\rm C},\,V_{\rm DH},\,V_{\rm H}$ or $V_q$),
be at the nucleus with coordinate $r=0$ and the normalised probability that the electron velocity be $v$, probability given by the distribution function $f_q(v)$ ($q=1$ for C, DH and H, where $f_{_{q=1}}(v)\equiv f_{_{\rm MB}}(v)$ is the normalized MB distribution of electrons). This distribution must be appropriate to the screening potential used.

We define the pure Coulomb nuclear electron capture rate, averaged over a MB distribution, as
\begin{equation}
{\cal R}_{_{\rm C}}(T)=\int\limits_0\limits^\infty(\sigma_{\rm e}\,v)\,F_{_{\rm C}}\,f_{_{\rm MB}}(v)\,4\,\pi\,v^2\,dv ,\label{rc}
\end{equation}
where
\begin{equation}
\sigma_{\rm e}={G^2\over\pi\,(\hbar\,c)^4}\,{c\over v}\,\Big(W_0+W\Big)^2\,\chi \ ,
\end{equation}
is the nuclear electron capture cross section \cite{Bahcall1,Bahcall2} with $G$ the Fermi constant,
$W_0$ the nuclear energy release for one electron with total energy $W$, $\chi=C_{\rm V}^2\,\langle1\rangle^2+C_{\rm A}^2\,\langle\sigma\rangle^2$
the well-known reduced nuclear matrix element \cite{Bahcall1}.\\
The Fermi factor for Coulomb potential, given by $F_{_{\rm C}}(E)=2\,\pi\,\eta/(1-e^{-2\,\pi\,\eta})$  with $\eta=4/(a_0\,p)$ where $a_0$ is the Bohr radius and  $p=m_e\,v$ is the electron momentum, follows from the definition $F_{_{\rm C}}(E)=\lim_{r\to0}|\psi_{_{\rm C}}(r)/p\,r|^2$, where $\psi_{_{\rm C}}(r)$ is the wave function of the Schr\"odinger equation with the Coulomb potential.\\
The Hulth\'{e}n rate ${\cal R}_{_{\rm H}}$ has been evaluated by averaging over MB distribution and by substituting in
the Coulomb rate of Eq. (\ref{rc}) the Fermi factor $F_{_{\rm H}}(E)=\lim_{r\to0}|\psi_{\rm H}(r)/p\,r|^2$, where $\psi_{_{\rm H}}(r)$ is now the wave function of the Schr\"odinger equation with the Hulth\'{e}n potential
$V_{\rm H}(r)=-(Z\,e^2/R_{\rm DH})\,e^{-2\,r/R_{\rm DH}}\,(1-e^{-2\,r/R_{\rm DH}})^{-1}$
and, in this case, the Fermi factor can be obtained analytically \cite{Durand,Yoon}.\\
Finally, the non extensive rate ${\cal R}_q$, can be obtained by substituting in Eq. (\ref{rc}) the factor
$F_{_{\rm C}}(E)$ with the Fermi factor $F_q(E)=\lim_{r\to0}|\psi_q(r)/p\,r|^2$
where $\psi_q(r)$ can be obtained as a numerical solution of the Schr\"oedinger equation with the modified
DH potential \cite{Quarati}
\begin{equation}
V_q(r)=-{Z\,e^2\over r}\left[1-(1-q)\,\delta_q\,r\right]^{1/1-q} \ ,
\end{equation}
with $\delta_q=1/[(2-q)\,\xi]$.\\
Consistently with the derivation of the MDH potential in the definition of ${\cal R}_q$ we must insert in place of the MB distribution the normalised non extensive distribution $f_q(v)$ \cite{Quarati} defined in
\begin{equation}
f_q(v)=B_q\,\left({m_e\over2\,\pi\,k\,T}\right)^{3/2}\exp_q\left(-{m_e\,v^2 \over 2\,k\,T}\right) \ ,\label{fq}
\end{equation}
with $\exp_q(x)=\left[1+(1-q)\,x\right]_+^{1/1-q}$ the $q$-exponential ($[x]_+=x\,\theta(x)$ where $\theta(x)$ is the Heaviside step function) and for $q<1$
\begin{equation}
B_q=\sqrt{1-q}\,{5-3\,q\over2}\,{3-q\over2}\,
{\Gamma\left({1\over2}+{1\over1-q}\right)\over\Gamma\left({1\over1-q}\right)} \ .
\end{equation}
For $q=1$ formula (\ref{fq}) reduces to the MB distribution.\\
The integral for the rate ${\cal R}_q$, when $q<1$, is performed over the real interval $[0,\,v_{\rm cut}]$ with $v_{\rm cut}=\sqrt{2\,k\,T/[(1-q)\,m_e]}<1$ which defines a cut-off condition in the velocity space. When $q\to1$ the rate ${\cal R}_q$ reduces to DH rate ${\cal R}_{_{\rm DH}}$.

\section{Results}

We pose our attention to the case of solar EC by $^7$Be. Results can be easily extended to any $(A,\,Z)$ nucleus (calculations are in progress).\\ In figure 1, we report, for several value of $\xi$ the quantity
\begin{equation}
\Delta P_{_{\rm X}}(E)={F_{_{\rm C}}(E)-F_{_{\rm X}}(E)\over F_{_{\rm C}}(E)} \,
\end{equation}
where X\,$ = $\,H, DH and $q$ (with $q=0.95,\,0.85,\,0.75$) that represents the percentage variation of probability density at $r=0$ compared to the probability density when the screening potential is a pure Coulomb potential.
\begin{center}
\begin{figure}[h]
  \hspace{20mm}\resizebox{120mm}{!}{\includegraphics{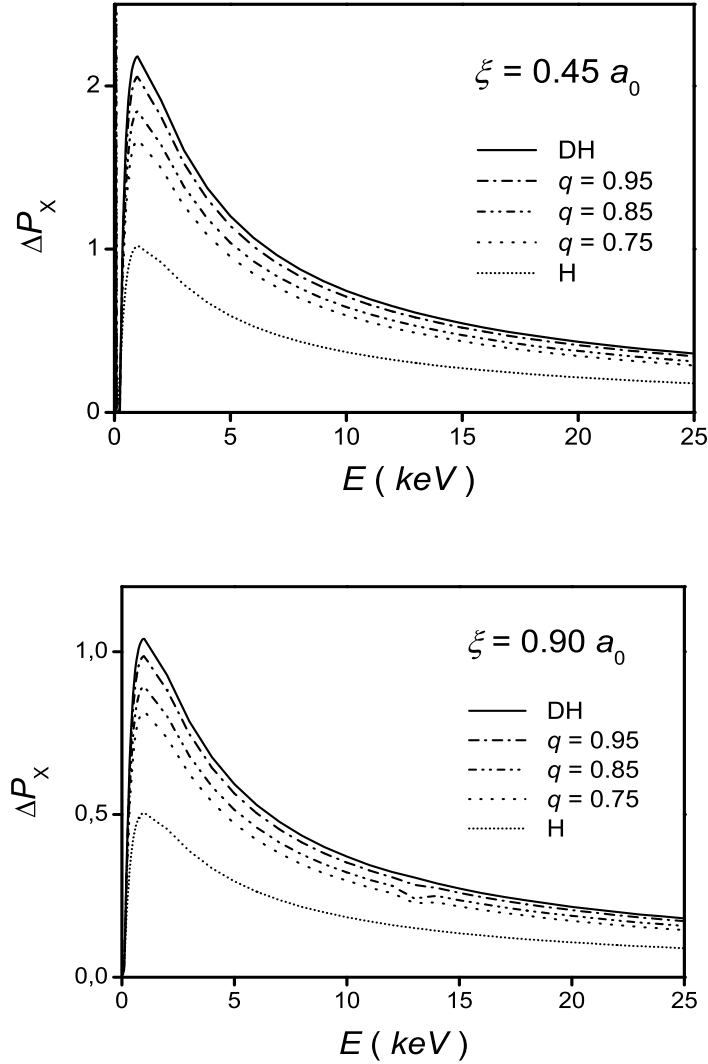}}
  \caption{Relative variation $\Delta P_{_{\rm X}}(E)$ of the Fermi factor $F_{_{\rm X}}(E)$ over the Coulomb Fermi factor $F_{_{\rm C}}(E)$ for $\xi=0.45\,a_0$ and $\xi=0.90\,a_0$ and energy $0\,keV < E< 25\,keV$.}
\end{figure}
\end{center}
Bahcall and Moeller \cite{Bahcall3} and Gruzinov and Bahcall \cite{Gruzinov} have found that, for $R_{_{\rm DH}}>0.4\,a_0$ and $Z=4$, the quantity $\Delta P_{_{\rm DH}}(E)$ is less than $1$ percent, for $0.3\,a_0\leq R_{_{\rm DH}}\leq0.4\,a_0$ about $2$ percent and have concluded that plasma screening is unimportant for capture from continuum. Our calculation of DH electron density at $r=0$ (which corresponds to the case of $q=1$) agrees with their results. However, in \cite{Bahcall3} and \cite{Gruzinov} the above authors have calculated the rate averaging over MB distribution, in the frame of a global thermodynamical equilibrium, therefore they have found a negligible screening effect over Coulomb rate and a negligible effect of fluctuations and correlations over DH rate. Although electron density at $r=0$ due to $V_q(r)$ is smaller than Coulomb density, in the velocity space the probability density in the low momentum region is greater than MB because the continuum electron distribution $f_q(v)$ we use in this work privileges low momentum electrons. Therefore, screening may be important in continuum EC rate. This can be seen in Table 1, where we report the calculated deviations of the rate ${\cal R}_{_{\rm X}}(T)$ respect to ${\cal R}_{_{\rm C}}(T)$, by means of the function
\begin{equation}
\Delta\Omega_{_{\rm X}}(T)={{\cal R}_{_{\rm C}}(T)-{\cal R}_{_{\rm X}}(T)\over{\cal R}_{_{\rm C}}(T)} \ ,
\end{equation}
at the value of $k\,T=1.27\,keV$ (where EC by $^7$Be takes place), for the three values of $q$ of figure 1.\\

\begin{table}
\caption{Table 1. Relative variation $\Delta\Omega_{_{\rm X}}(T)$ of the rate ${\cal R}_{_{\rm X}}(T)$ over the
Coulomb rate ${\cal R}_{_{\rm C}}(T)$ for several values of $\xi$, at the temperature $k\,T= 1.27\,keV$, for the three values of $q$ of figure 1.}
\begin{indented}
\item[]
\begin{tabular}{lccccc}\br
 & \ H \ \ & \ \ DH \ & \ $q=0.95$ \ & \ $q=0.85$ \ & \ $q=0.75$ \\
\mr
$\xi=0.30\,a_0$ \ & 2.24\% \ \ & 0.58\% & -2.92\% & -9.45\% & -15.44\% \\
$\xi=0.45\,a_0$ \ & 1.82\% \ \ & 0.69\% & -2.65\% & -9.03\% & -15.06\% \\
$\xi=0.60\,a_0$ \ & 1.62\% \ \ & 0.45\% & -2.91\% & -9.36\% & -15.48\% \\
$\xi=0.75\,a_0$ \ & 1.52\% \ \ & 0.23\% & -3.13\% & -9.58\% & -15.70\% \\
$\xi=0.90\,a_0$ \ & 1.45\% \ \ & 0.08\% & -3.28\% & -9.73\% & -15.82\% \\
\br
\end{tabular}
\end{indented}
\end{table}

For any $q<1$, ${\cal R}_q>{\cal R}_{_{\rm C}}$. We have verified that deviations depend very smoothly on $k\,T$ except for $\xi\leq0.45\,a_0$ and depend very strongly on $q$.\\
The value of $q$ for EC by $^7$Be in solar plasma can be derived from the expression that links $q$  to fluctuation of $1/R_{_{\rm DH}}$ \cite{Quarati}. By using the equation of state for $q$-nonextensive systems \cite{Plastino} we obtain for $0<q<1$
\begin{equation}
\sqrt{1-q}={\Delta N_{_{\rm DH}}\over N_{_{\rm DH}}}={1\over\sqrt{N_{_{\rm DH}}}}\,{1\over\sqrt{1+(1-q)\,K}} \ ,
\end{equation}
where $K={3\over2}\,N_{_{\rm DH}}\,\ln\left[0.211\cdot(0.45\,a_0)^2\right]-\ln(N_{_{\rm DH}}!)$.\\
In the solar core, where the average electron density is $n_e=9.1\,a_0^{-3}$, $N_{_{\rm DH}}$, the number of particles inside the Debye sphere, is about $4$, we can derive $q=0.86$. It is more safe to consider a range of values of $q$ between $0.84$ and $0.88$. At $k\,T=1.27\,keV$ and $\xi=0.45\,a_0$ the calculated ${\cal R}_q(T)$
is estimated to be about $7$ -- $10$ percent over standard DH ($q=1$) estimate.\\
Let us consider the $^7$Be -- p fusion, reaction producing $^8$B and, as a consequence, $^8$B neutrinos, in competition with $^7$Be electron capture \cite{Lissia}. We have verified that the effect of the MDH potential over its rate is negligible. In fact, correction to $F_{_{\rm C}}$ is effective only at relative $^7$Be -- p energies lower than $2.4\,keV$ where fusion cross section has a negligible value because its most effective energy is at $18\,keV$. Therefore, if EC rate of $^7$Be increases over its standard evaluation of a given percentage, $^7$Be increases its destruction while the neutrino flux from $^7$Be does not change because the $^7$Be density decreases. However, the $^8$B flux should diminish of the same percentage. This behaviour is in line with what is found in experiments \cite{Haxton}.

\section{Conclusions}

We have calculated the EC rate ${\cal R}_q(T)$ for electrons of the continuum spectrum, with the use of the MDH screening potential $V_q(r)$ (derived in \cite{Quarati}). The value of the parameter $q$ has been estimated from its relation with the fluctuation of the number of particles contained in the Debye sphere, in the solar core. It is reasonable to take the range of values $0.84<q<0.88$. We have calculated the rate averaging over the electron non-Gaussian $q$-distribution $f_q(v)$.\\
Considering the EC by $^7$Be at $\xi=0.45\,a_0$ and $k\,T=1.27\,keV$ we have evaluated an increase of the capture rate ${\cal R}_q(T)$ of $7$ -- $10$ percent over standard DH rate ${\cal R}_{_{\rm DH}}(T)$ that is the $0.69$ percent smaller, at the same conditions, than Coulomb rate ${\cal R}_{_{\rm C}}(T)$. Of course, a smaller value of $q$ should imply a much greater enhancement of EC rate over DH one. The main consequence concerns the calculated neutrino $^8$B flux that decreases of the same percentage respect to its evaluation with standard DH screening, while $^7$Be flux remains unchanged.


\end{document}